\documentclass[conference]{IEEEtran}
\IEEEoverridecommandlockouts
\usepackage{cite}
\usepackage{amsmath,amssymb,amsfonts}
\usepackage{algorithmic}
\usepackage{graphicx}
\usepackage{textcomp}
\usepackage{xcolor}
\usepackage{siunitx}  
\usepackage{listings}

\def\BibTeX{{\rm B\kern-.05em{\sc i\kern-.025em b}\kern-.08em
    T\kern-.1667em\lower.7ex\hbox{E}\kern-.125emX}}
\begin{document}

\title{Neural 5G Indoor Localization with IMU Supervision}
\author{\IEEEauthorblockN{Aleksandr Ermolov\IEEEauthorrefmark{1}, Shreya Kadambi\IEEEauthorrefmark{2}, Maximilian Arnold\IEEEauthorrefmark{1}, Mohammed Hirzallah\IEEEauthorrefmark{2},\\Roohollah Amiri\IEEEauthorrefmark{2},
Deepak Singh Mahendar Singh\IEEEauthorrefmark{2}, Srinivas Yerramalli\IEEEauthorrefmark{2},\\Daniel Dijkman\IEEEauthorrefmark{1}, Fatih Porikli\IEEEauthorrefmark{2}, Taesang Yoo\IEEEauthorrefmark{2}, Bence Major\IEEEauthorrefmark{1}}
\IEEEauthorblockA{
\IEEEauthorrefmark{1}Qualcomm Technologies Netherlands B.V.,
\IEEEauthorrefmark{2}Qualcomm Technologies Inc.\\
\{aermolov,skadambi,marnold,bence\}@qti.qualcomm.com}
}

\maketitle

\begin{abstract}
Radio signals are well suited for user localization because they are ubiquitous, can operate in the dark and maintain privacy. Many prior works learn mappings between channel state information (CSI) and position fully-supervised. However, that approach relies on position labels which are very expensive to acquire. In this work, this requirement is relaxed by using pseudo-labels during deployment, which are calculated from an inertial measurement unit (IMU). We propose practical algorithms for IMU double integration and training of the localization system. We show decimeter-level accuracy on simulated and challenging real data of 5G measurements. Our IMU-supervised method performs similarly to fully-supervised, but requires much less effort to deploy.
\end{abstract}


\section{Introduction}

This work is focused on the localization task, i.e. locating the user equipment (UE) in some specific indoor environment from available measurements. Such methods can be classified based on input modalities: light detection and ranging (Lidar) measurements, visual observations, inertial measurement unit (IMU), radio signal. The latter is of particular interest \cite{del2017survey} due to the widespread use of the technology and its privacy-preserving properties.

Radio signal localization is a well-known task, where time of flight of radio signals from the transmitter to the receiver is measured to derive the user location. This requires either synchronization at the network side (for downlink or uplink time difference of arrival, i.e., DL-TDOA and UL-TDOA), bi-directional signaling used for round-trip time-based measurements (i.e., RTT) or angle-based estimates (Angle of Departure, AoD or Angle of Arrival, AoA) if multiple antennas are available. Fundamentally, all these methods assume the presence of Line-of-Sight (LoS) in the observed channel. If excess of measurements are present, outlier rejection methods can be employed to reduce the impact of non-Line of Sight (NLoS) measurements on the final positioning accuracy.

\begin{figure}[t]
\centerline{\includegraphics[width=8cm]{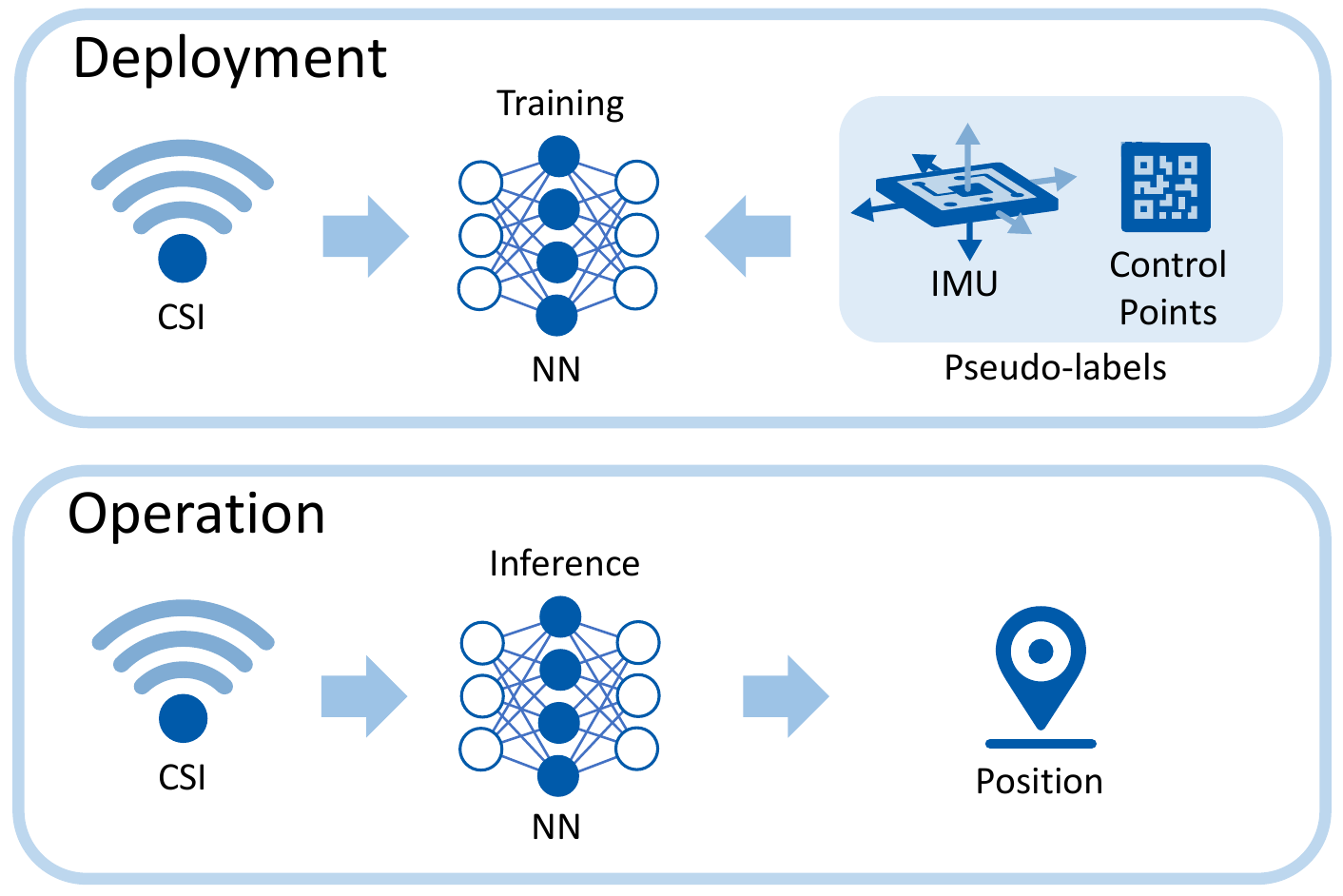}}
\caption{Two stages of our method. Note that during operation the position prediction does not depend on past or future steps, only observed CSI is used.}
\label{fig:deploy}
\end{figure}

In recent literature, supervised machine learning methods \cite{savic2015fingerprinting,khalajmehrabadi2017modern,liu2018autloc,vieira2017deep,arnold2019novel} have been proposed for dealing with challenging LoS conditions. These methods train an algorithm to learn a mapping from the observed \emph{Channel State Information (CSI)} of multiple cells to the coordinates at which the CSI is measured. Training such algorithms generally requires collecting CSI from the target environment densely along with the user location. Acquiring such precise positioning labels requires a careful measuring process, which makes this approach somewhat impractical to deploy. This limitation has given rise to semi-supervised and unsupervised learning methods \cite{cc_studer2018,karmanov2021wicluster,ghazvinian2021modality,gill2021three,kadambi2022neural,arnold2022benchmarking}.

The availability of CSI, which contains implicit information about the radio environment and spatial geometry of the users has enabled \emph{Channel Charting} methods \cite{cc_studer2018} to learn a low dimensional representation from CSI that still preserves the time-varying and space-varying components in an unsupervised fashion. However, these methods still require samples with known labels to accurately predict the location of the user from learned low-dimensional representation.

IMU measurements do not depend on expensive or privacy-breaching sensors, but positions, obtained from IMU integration, are drifting over time, accumulating the error. There are several very recent works that suggest to combine radio and IMU modalities \cite{yang2023novel,yu2022multi,zhang2023csi,long2021csi}. Our work continues this line of research.
In this proposal, we significantly relax the requirement of dense position labels, by replacing the manual position measurement process with automatically obtained “pseudo-labels”. Such pseudo-labels could be acquired using sensors such as a vision camera or Lidar, but such system generally have a higher cost and are not privacy-preserving. Our solution acquires these pseudo-labels from IMU measurements.

In summary, our contributions are the following:
\begin{itemize}
  \item We propose a neural network (NN) based positioning system trained on pseudo-labels, obtained from IMU double integration.
  \item To improve pseudo-labels accuracy, we propose trajectory fitting algorithm, which combines IMU dead-reckoning with limited control point information.
  \item We demonstrate that IMU trajectory fitting can be iteratively refined using CSI information further improving the accuracy.
  \item We prototype our solution on a simulated dataset and then verify the takeaways with a real data collection setup. The model shows decimeter-level accuracy while using only single control point and noisy IMU measurements during deployment.
\end{itemize}

\section{Method}

\begin{figure}[htbp]
\centerline{\includegraphics[width=9cm]{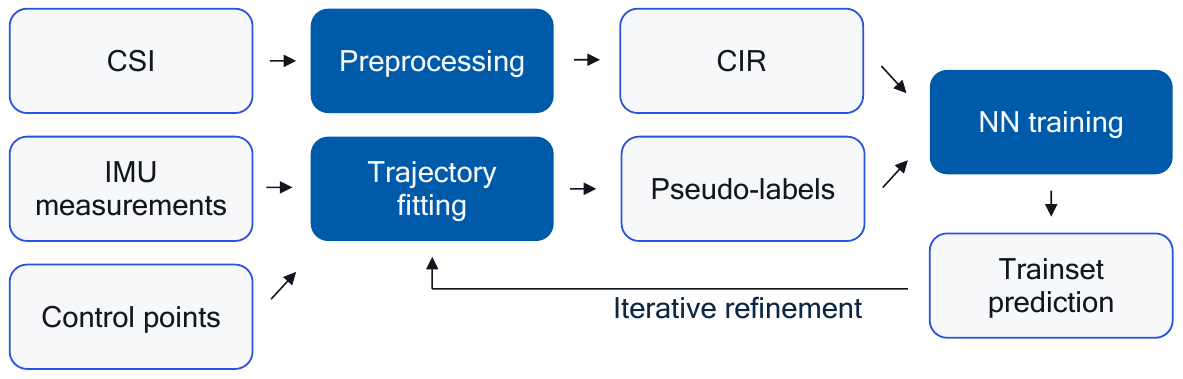}}
\caption{Training pipeline.}
\label{fig:pipeline}
\end{figure}

The positioning solution consists of two stages: deployment and operation (Fig.~\ref{fig:deploy}). At the deployment stage, the UE moves throughout the environment, with the CSI and IMU measurements collected. A few prominent points on the ground are marked with fiducials. At these points, the coordinates and corresponding velocities are computed using a privacy-preserving floor-facing camera. These points are the \emph{control points}. An automated process then computes the pseudo-label coordinates for the remaining points, by finding a most likely trajectory, derived from IMU measurements passing through control points. The last step of deployment is training a neural network for each time step to predict from the CSI of that time step the pseudo-label position.
At operation time, the CSI is fed to the neural network, which estimates the UE position. Note that the control points with floor-facing camera and IMU measurements are only used during deployment, and \emph{only} CSI is required during operation.

Our training pipeline consists of several stages (Fig.~\ref{fig:pipeline}). First, we preprocess the raw CSI information. Next, we compute pseudo-labels from IMU measurements and control points. After that, we train the NN using obtained information. Finally, we refine the pseudo-labels and retrain the NN.

\subsection{Preprocessing}
For a real data collection setup, channel state information is measured on the uplink at the transmission-reception-points (TRPs). For all our results we are using 5G numerology for the industry case, where we measure using a bandwidth of 100MHz, a subcarrier spacing of 30kHz the channel frequency response at the pilot locations. Thus only each fourth subtone is occupied. The channel impulse response is created by calculating the inverse Fourier transformation on these down-sampled subtones.

As mentioned in the section on the data set, since UE and TRPs are not synchronized we compensate for the drift in the pre-processing. The group delay offset is correct for when the measurements are collected. 
\\
\subsubsection{LoS peak Alignment}
For real data, inspired from classical time-difference-of-arrival estimation techniques, one can remove the arbitrary time/phase shift of the transmit starting time, via LoS peak alignment.
To compensate for the clock offset drift we align the line of sight peaks of a reference antenna. A showerhead TRP$_{0}$ provides a good LoS path across most of the trajectory, antenna 0 of TRP$_{0}$ is chosen as a reference.

\begin{figure}[htbp]
\centerline{\includegraphics[width=8cm]{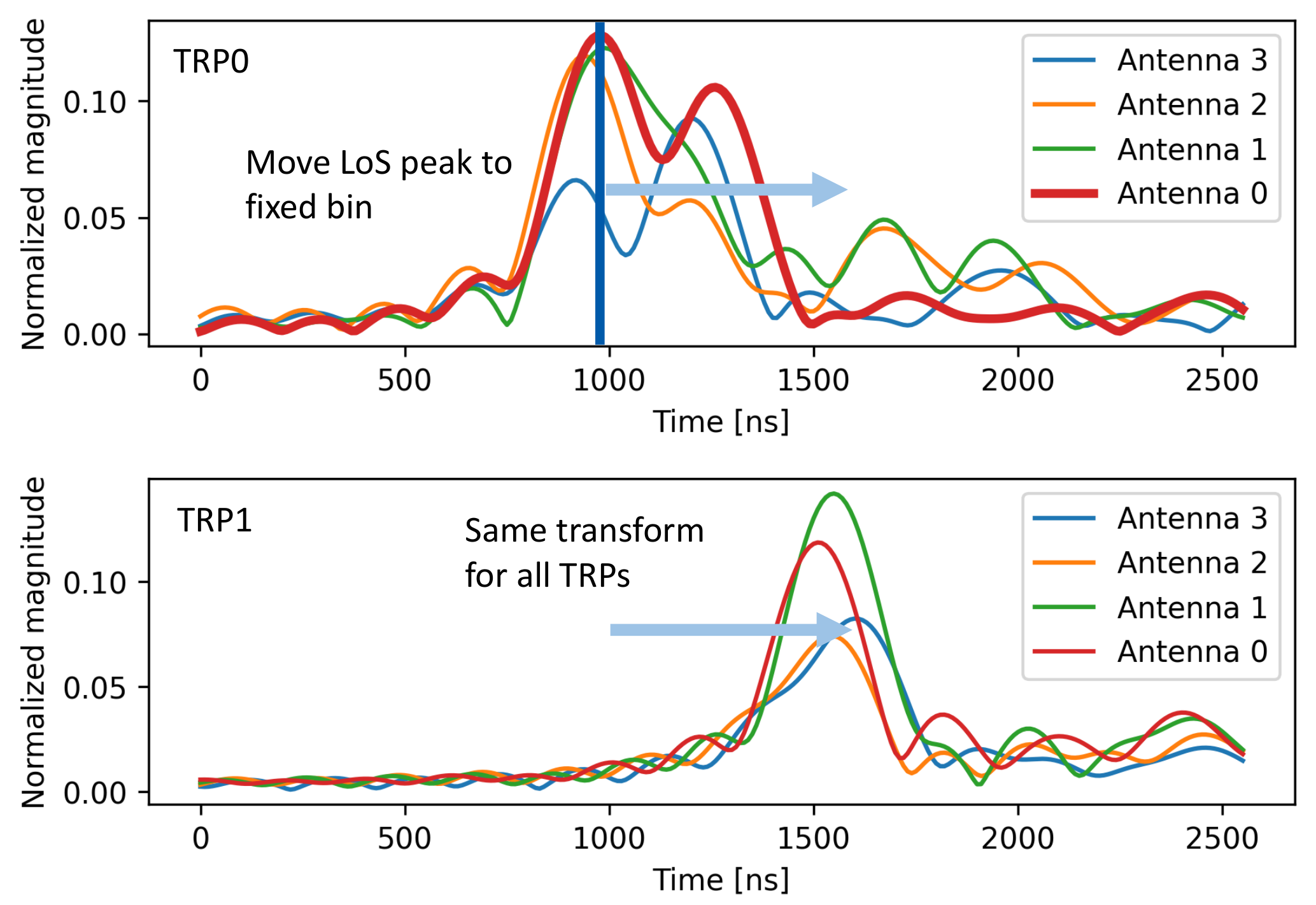}}
\caption{LoS alignment procedure. The LoS peak is detected and CIR is rolled to align the peak to a specific bin. The same shift applies to all TRPs and antennas.}
\label{fig:los-align}
\end{figure}

For each CSI packet, the first measured peak of the reference antenna \(T_0\) is shifted to arrive at a fixed time step \(T_\text{bin}\). If the peak is not found, dataset median peak is used as a fallback. The effect of peak alignment can be seen in the Fig.~\ref{fig:los-align}. For the rest of the TRPs and antennas, channel impulse response (CIR) is shifted by the same amount $T_{\text{bin}} - T_0$.

\subsubsection{Processing in time domain}
We obtain the SNR information for each antenna by computing the power on the pilot symbols and noise on the non-reference symbols.
Before training, we remove samples with SNR below a threshold and normalize the time domain response per antenna.

\subsection{Forward-backward algorithm}

\begin{figure}[htbp]
\centerline{\includegraphics[width=9cm]{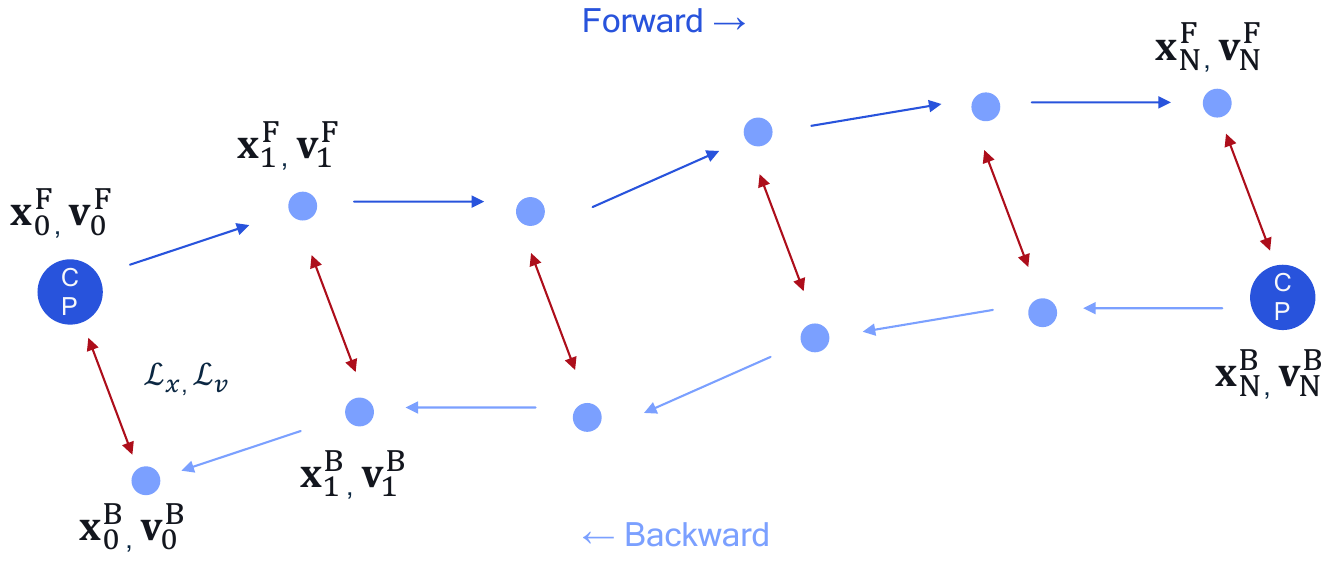}}
\caption{Forward-backward algorithm. The scheme depicts the trajectory fitting based on misalignment of the positions obtained with forward and backward integration passes. Respectively, each position of the trajectory is represented with two points. We optimize the correction of the acceleration in each position pulling these two points to each other (shown with red arrow).}
\label{fig:fwd-bwd}
\end{figure}

IMU data includes linear acceleration, angular velocity and device orientation. We assume that the sensor orientation is calibrated and we use global acceleration, obtained from it, to recover positions.

The key idea of the algorithm is to fit the trajectory of the UE between two control points using IMU data for each step. The algorithm takes as input: $\Delta t_i$ — time difference between steps, $\mathbf{a}_i^\text{IMU}$— noisy acceleration obtained from IMU, positions and velocities in control points ($\mathbf{x}_0^\text{F}, \mathbf{v}_0^\text{F}, \mathbf{x}_N^\text{B}, \mathbf{v}_N^\text{B}$).

The trajectory positions can be recovered with double integration (dead-reckoning):
\begin{equation}
\begin{split}
\mathbf{v}_n^\text{IMU} = \mathbf{v}_{n-1}^\text{IMU} + \mathbf{a}_n^\text{IMU} \Delta t_n,\\
\mathbf{x}_n^\text{IMU} = \mathbf{x}_{n-1}^\text{IMU} + \mathbf{v}_n^\text{IMU} \Delta t_n
\end{split}
\end{equation}
For each step, we train a correction parameter $\mathbf{a}_n^\text{cor}$ that compensates the noise drift. Resulting acceleration $\tilde{\mathbf{a}_n} = \mathbf{a}_n^\text{IMU} + \mathbf{a}_n^\text{cor}$.
In this case, forward and backward integration:
\begin{equation}
\mathbf{v}_n^\text{F} = \mathbf{v}_{n-1}^\text{F} + \tilde{\mathbf{a}_n} \Delta {t}_n, \; \mathbf{x}_n^\text{F} = \mathbf{x}_{n-1}^\text{F} + \mathbf{v}_n^\text{F} \Delta t_n
\end{equation}
\begin{equation}
\mathbf{v}_{n-1}^\text{B} = \mathbf{v}_n^\text{B} - \tilde{\mathbf{a}_n} \Delta t_n, \; \mathbf{x}_{n-1}^\text{B} = \mathbf{x}_n^\text{B} - \mathbf{v}_n^\text{B} \Delta t_n
\end{equation}
Finally, our optimization losses:
\begin{equation}
\begin{split}
\mathcal{L}_x = \sum_i \Vert \mathbf{x}_i^\text{F} - \mathbf{x}_i^\text{B} \Vert_2^2 \\
\mathcal{L}_v = \sum_i \Vert \mathbf{v}_i^\text{F} - \mathbf{v}_i^\text{B} \Vert_2^2 \\
\mathcal{L}_\text{reg} = \sum_i \Vert \mathbf{a}_i^\text{cor} \Vert_2^2
\end{split}
\end{equation}

\begin{figure}[htbp]
\centerline{\includegraphics[width=7.5cm]{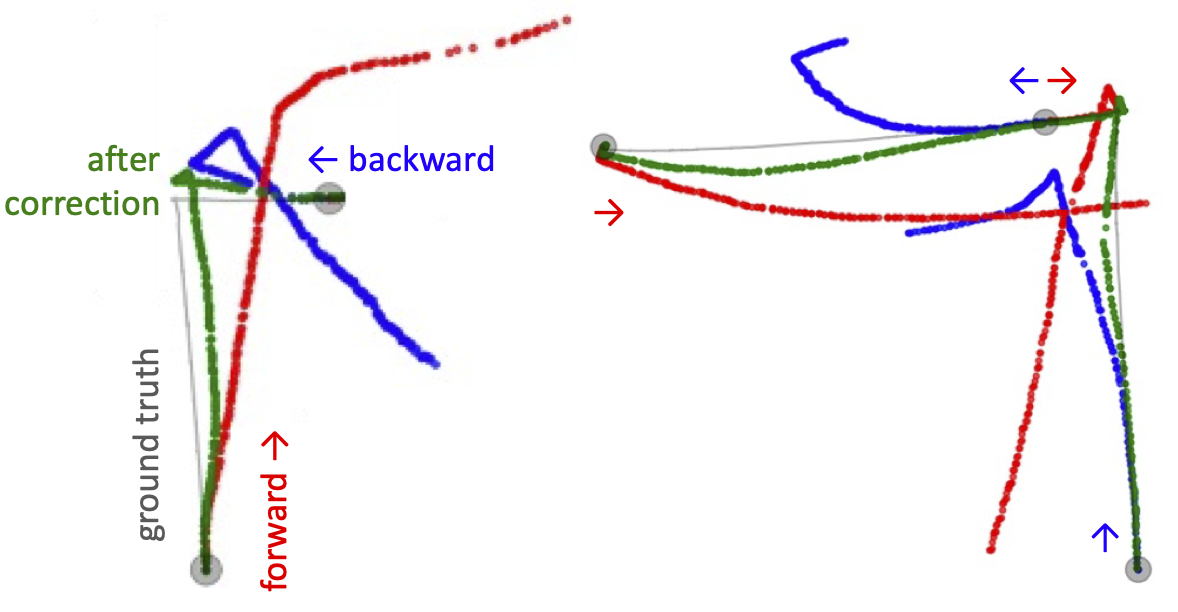}}
\caption{Simulated trajectories. Gray circles show control points, which act as starting points for integration. Red and blue trajectories are obtained with forward and backward integration. They quickly diverge from the ground truth due to the error accumulation. Green trajectory is the result of the forward-backward algorithm, it corrects the error pushing the trajectory closer to the ground truth.}
\label{fig:correction}
\end{figure}

$\mathcal{L}_x$ and $\mathcal{L}_v$ optimize $\mathbf{a}_n^\text{cor}$ such that forward and backward integrations produce close trajectories. At the same time, $\mathcal{L}_\text{reg}$ pushes the  correction to be as small as possible, assuming that the IMU noise is zero-centered. The scheme of the algorithm is depicted on Fig.~\ref{fig:fwd-bwd}. An example of forward integration, backward integration and correction is shown on Fig.~\ref{fig:correction}. The pseudo-labels are generated as the average position of the forward and backward passes after correction.

\subsection{Iterative refinement algorithm}

\begin{figure}[htbp]
\centerline{\includegraphics[width=4cm]{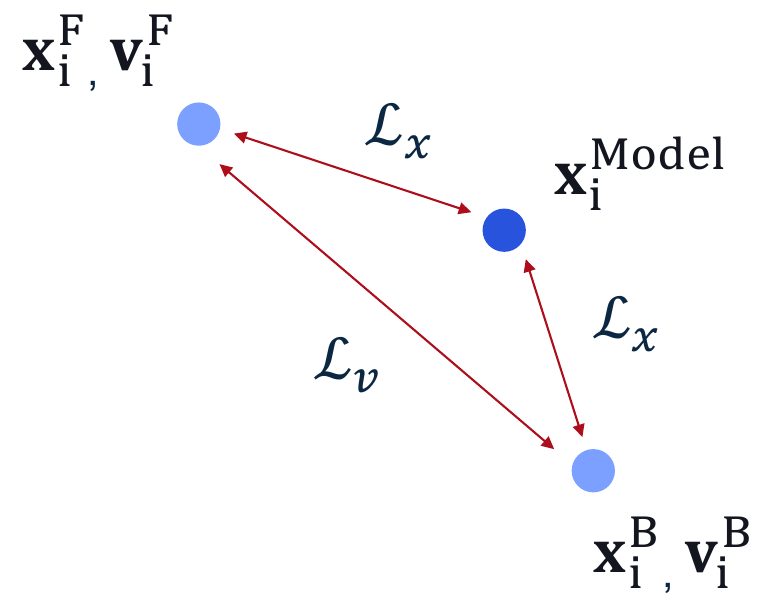}}
\caption{Iterative refinement losses. The algorithm extends the scheme~\ref{fig:fwd-bwd} with model prediction point $\mathbf{x}_i^\text{Model}$, and points, obtained from integration, are pulled to it.}
\label{fig:refine}
\end{figure}

In some cases, the accuracy of the trained NN can surpass the accuracy of IMU pseudo-labels. E.g., it is a common situation when the density of train samples is high and the network is trained on many close-by samples with different label errors. In such cases, the network can generalize well reducing these errors. As a result, the network can be used to refine IMU trajectory fitting.
The complete algorithm:
\begin{enumerate}
\item Create pseudo-labels with forward-backward algorithm.
\item Initialise the network randomly, train on pseudo-labels.
\item Predict positions for training set $\mathbf{x}_i^\text{Model}$ using the trained network.
\item Create pseudo-labels with forward-backward algorithm and predictions. In this case, the forward-backward scheme includes additional prediction point (positions from the previous step) and $\mathcal{L}_x$ is calculated between positions $\mathbf{x}_i^\text{F}, \mathbf{x}_i^\text{Model}$ and $\mathbf{x}_i^\text{B}, \mathbf{x}_i^\text{Model}$ (Fig.~\ref{fig:refine}).
\item Go to point 2 until convergence.
\end{enumerate}

\section{Experimental setup}
In this section we discuss our experimental setup including the datasets.  We start with a simulated environment, used for prototyping the solution. Next, we present real warehouse-like setup, used to verify our takeaways.

\begin{figure}[htbp]
\centerline{\includegraphics[width=6cm]{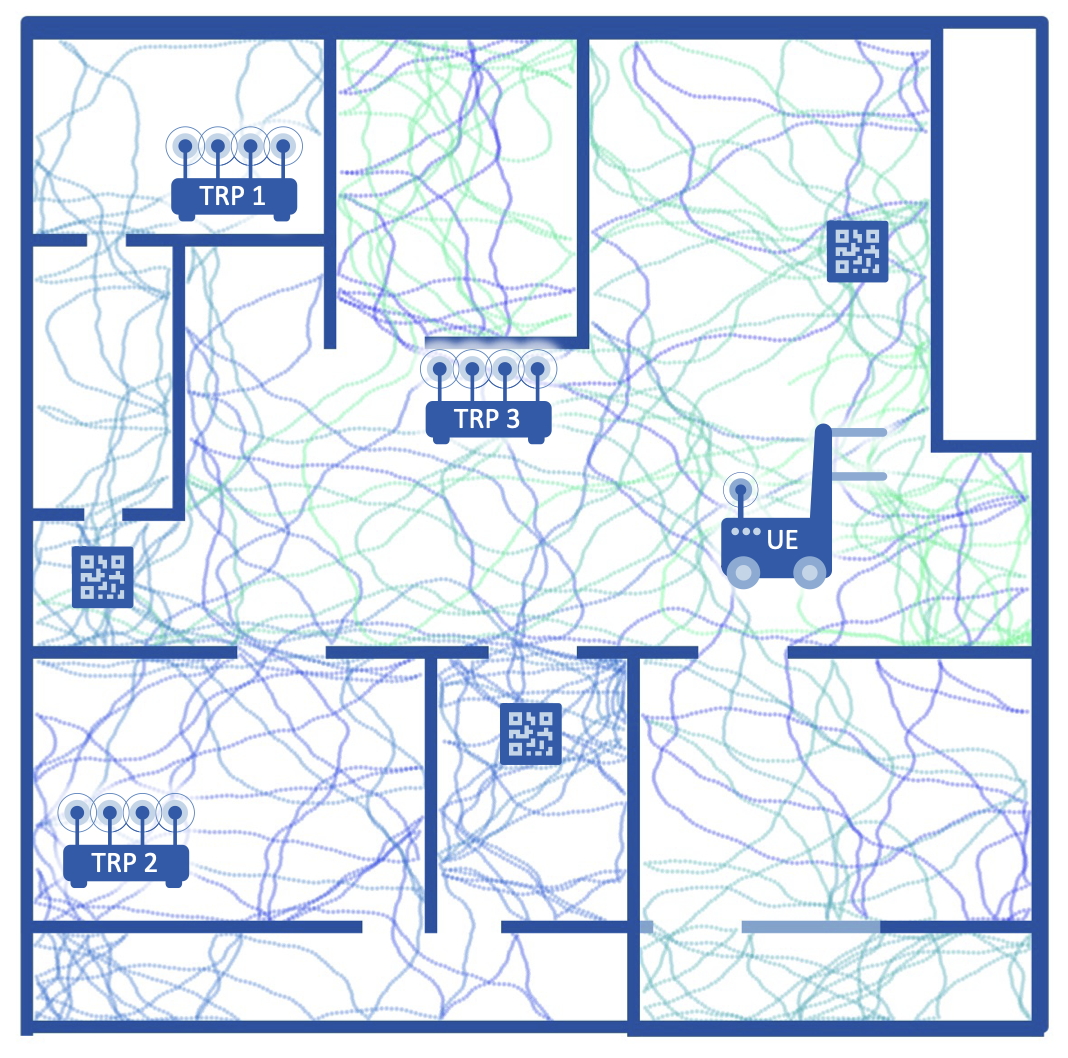}}
\caption{Data simulation. $\SI{30}{\metre} \times \SI{20}{\metre}$ environment with 3 TRPs and 3 control points (QR codes on the floor). Colored background curve shows the trajectory of the UE, used for the experiment.}
\label{fig:sim}
\end{figure}

\subsection{Simulated dataset}
We use an indoor floorplan $\SI{30}{\metre} \times \SI{20}{\metre}$ (Fig.~\ref{fig:sim}) to create a simulated dataset using Wireless-Insite by RemCom \cite{insite}. In order to maintain realistic propagation characteristics we setup walls with different material properties. All inner walls were made of dielectric with strong permittivity $\epsilon = 2.8$. Outer walls along with the floor and ceiling were made of concrete with a thickness of $\SI{30}{\centi\metre}$ and permittivity 
of $\epsilon =5.31$. The doors are mostly free space however, the outer wall doors are configured with a glass material with $\epsilon = 2.4$. The transmitter and receiver antennas are configured with omnidirectional half-wave dipoles. Three access points are configured as receivers at a height of $\SI{3}{\metre}$ while the UEs are located at $\SI{1}{\metre}$ height allowing only a subset of trajectories to lie in the LoS region. We currently run X3D ray tracing method of Wireless Insite allowing each ray to undergo up to 1 diffraction, 2 transmissions, and 4 reflections. Wireless-Insite output is further bandlimited to $\SI{100}{\mega\hertz}$ with subcarrier spacing at $\SI{30}{\kilo\hertz}$ and with $3264$ subcarriers. We generate trajectories with a random walk. We randomly split samples into train and test sets (90\% and 10\%). The total number of samples is $8982$ with a $\SI{20}{\centi\metre}$ step size.

\begin{figure}[htbp]
\centerline{\includegraphics[width=7cm]{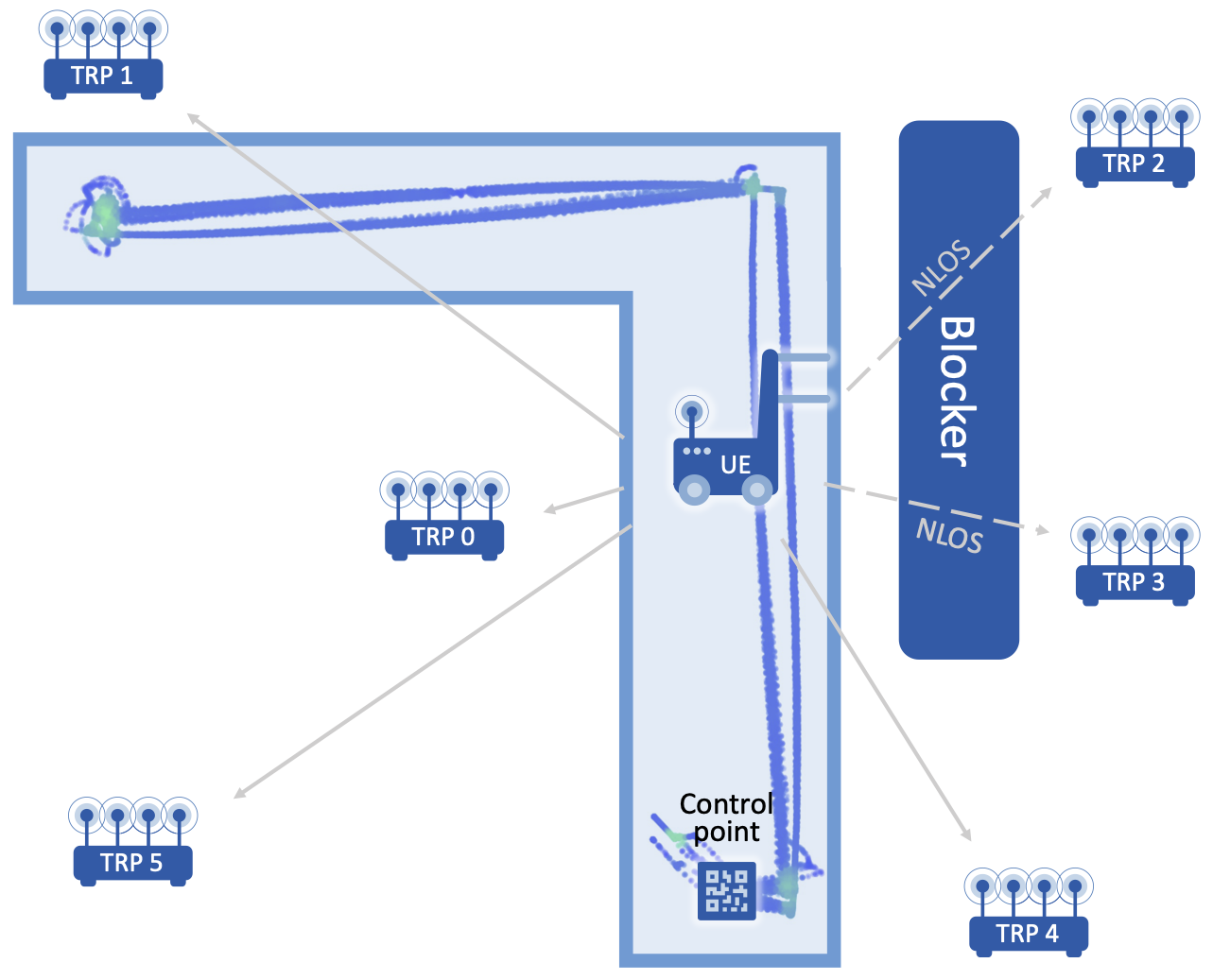}}
\caption{Warehouse dataset. The actual UE trajectory is shown on background, green color highlights regions with high points density.}
\label{fig:mwc23}
\end{figure}

\subsection{Warehouse dataset}
The network is deployed consisting of 6 TRPs surrounding a region of interest. Each TRP is deployed with 4 antennas. Two antennas in each TRP are vertically polarized and the other two are horizontally polarized respectively. The antenna for $\text{TRP}_0$ is a showerhead antenna while others are directional. The positions of the TRPs are known. Signals from all TRPs and all antenna ports are transmitted simultaneously. Similarly to simulation, the measured CSI has 3264 tones with a sub-carrier spacing of \SI{30}{\kilo\hertz}. Ground truth positions are obtained from laser pointers and QR codes.

\emph{Floor-plan}: UE is allowed to traverse an L-shaped region of size $\SI{15}{\metre} \times \SI{12}{\metre}$ as shown in the Fig.~\ref{fig:mwc23}. A thin metal sheet blocker is placed on the side blocking the LoS from $\text{TRP}_2$ and TRP$_3$ while allowing the refraction from the edges to pass (Fig.~\ref{fig:mwc23}). TRPs are mounted very close to the ceiling of the warehouse at a height of $\approx{4.5 m}$. UE has a single antenna. The total number of samples is $39425$ for training and $2775$ for testing. Average step size is $\SI{3.4}{\centi\metre}$ with $\SI{160}{\milli\second}$ time difference. The data was collected over two days.

\subsection{IMU data}

We simulate IMU measurements for our experiments. First, we calculate ground truth velocity and acceleration using coordinates and timestamps from datasets. Next, we input this acceleration into imuSensor system (part of Sensor Models, Navigation Toolbox, MATLAB), the configuration is listed in Tab.~\ref{tab:imu_spec}. The simulator outputs noisy acceleration, similar to smartphone-grade IMU\footnote{https://www.bosch-sensortec.com/products/motion-sensors/imus/bmi270/} measurements.

\subsection{Control points}
Control point is a point in which we can measure the position and velocity with high precision. In real environment it can be implemented with a floor fiducial marker and floor-facing camera on the UE. Radius defines the area and actual positions covered with the control point, we use $\SI{20}{\centi\metre}$ in all experiments (if not otherwise specified). We use $3$ control points in Simulated dataset (Fig.~\ref{fig:sim}), and $1$ in Warehouse (Fig.~\ref{fig:mwc23}). Additionally, we use random control points in the ablation experiments \ref{sec:cpoints}.

\subsection{Implementation details}
The localization model is a fully-connected neural network. It consists of three layers with $1024$, $512$ and $3$ output size. We use hyperbolic tangent activation functions. Input data, represented with first 256 bins of CIR for each antenna and each TRP, flattened into one vector.

We use Adam optimizer with learning rate $0.0001$ with $10\times$ drop for last 50 epochs. We use batch size $256$. We train for $600$ epochs for supervised baseline, $100$ epochs for the IMU-supervised experiments, $300$ epochs for control points number ablations, $100$, $200$, $300$ and $400$ epochs for iterative refinement with respect to each iteration. Using fewer epochs for the noisy ground truth provides some regularization to prevent overfitting to label imperfections. The localization is trained with smooth L1 loss for predicted positions. For Warehouse dataset, we use Kalman smoother for positions during validation with default\footnote{https://pykalman.github.io/} configuration.

We augment our data during training to improve generalization. In case of pseudo-labels, we add a uniform noise from range $(-5, 5)$ centimeters to positions. For Warehouse dataset, we add jittering to CIR with a random horizontal shift from a range of $(-7, 7)$ of total 256 bins.

For IMU trajectory fitting, we use stochastic gradient descent with learning rate $0.0001$. Correction parameters are initialized randomly from normal distribution with variance $\SI{0.0001}{\metre\per\second\squared}$. The total loss is a sum of three terms: $\mathcal{L} = \mathcal{L}_x + 10^3\mathcal{L}_v + 10^4\mathcal{L}_\text{reg}$, where the loss coefficients compensate magnitude difference of each term. The optimization is performed for $2000$ steps.

\section{Experimental Results}

We present the horizontal absolute error in centimeters of the model on the test set for two datasets. Our IMU-supervised experiments are performed for 30 random seeds, and IMU noise is independently sampled over seeds; tables show mean value with standard deviation.
For Simulated dataset, the area exceeds \SI{500}{\square\metre} inside a diverse environment, covered with only 3 TRPs, and our trajectory covers the area only partly. These challenges allow us to expand our findings to the real-case scenario. For Warehouse dataset, we have performed our experiments on measured 5G indoor data.

Tab.~\ref{tab:main_results} includes fully-supervised baseline. In this case, we use all the ground-truth labels during training. This baseline provides an \emph{upper bound} for our experiments, this accuracy can be obtained with perfect labels. Channel Charting baseline is provided for Warehouse dataset; for Simulated dataset this method was unable to learn the representation due to the sample sparsity. For this case, such type of methods requires additional heuristics \cite{karmanov2021wicluster}.

\begin{figure}[htbp]
\centerline{\includegraphics[width=7cm]{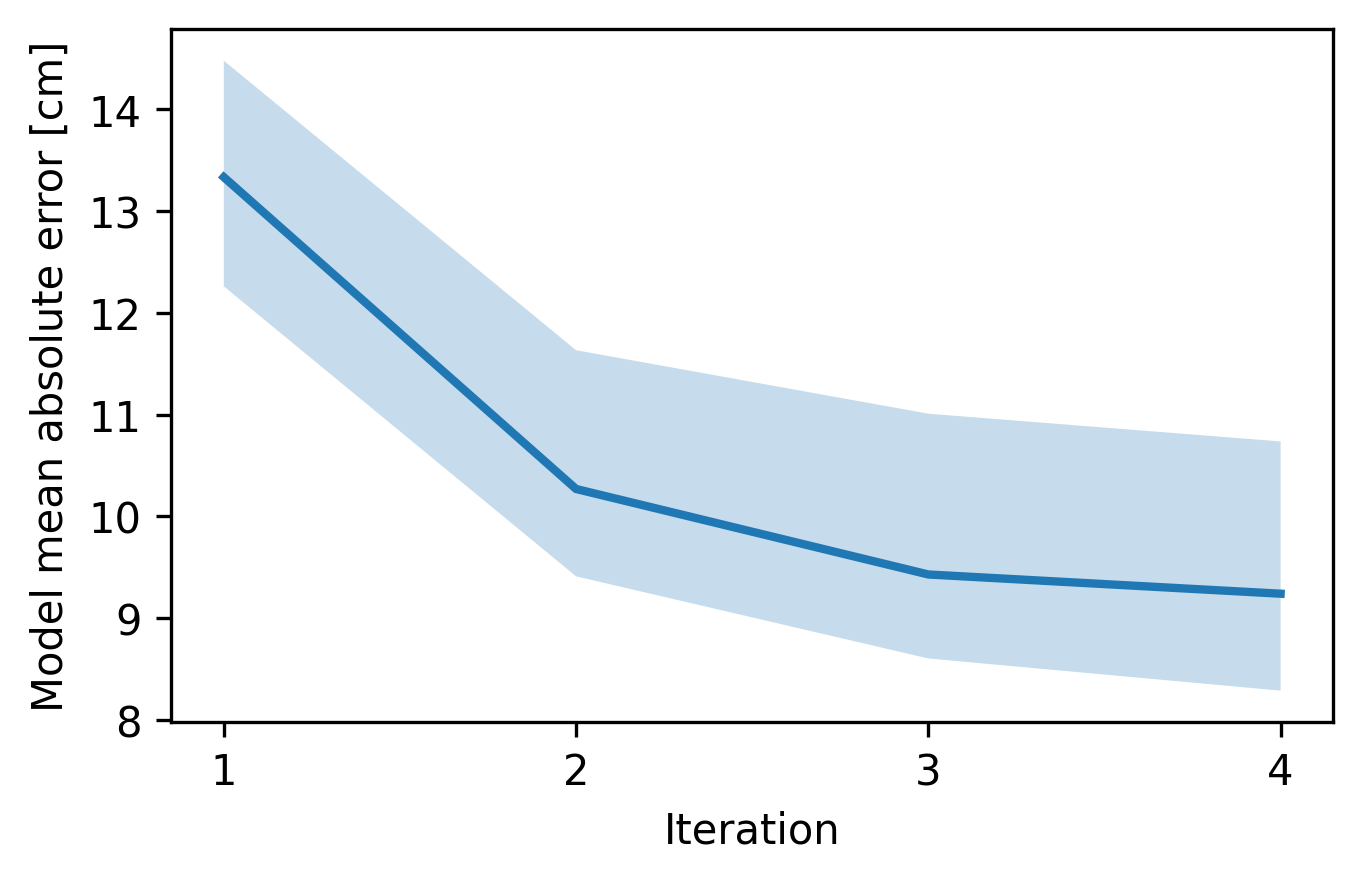}}
\caption{Model accuracy on each iteration of the refinement procedure. The light area is 0.1 and 0.9 quantiles, the line is median for 30 seeds.}
\label{fig:refine_plot}
\end{figure}

Results include two versions of our method, with and without iterative refinement (IR). We can see that our IMU-supervised model is very close to the fully-supervised upper bound, and iterative refinement further reduces the gap. \#CP column shows the number of control points for each method, the results confirm that our method reduces the deployment time effort.
For more experiments on control points we refer to ablation studies \ref{sec:cpoints}.
Additionally, Fig.~\ref{fig:refine_plot} demonstrates the performance of every iteration of the iterative refinement algorithm.

\begin{table}[htbp]
\def\arraystretch{1.3}
\caption{Horizontal absolute error [cm]}
\begin{center}
\begin{tabular}{|c|c|c|c|c|}
\hline
Method & \#CP & Mean & Median & 90\textsuperscript{th} perc.\\
\hline
\multicolumn{5}{c}{Simulated dataset} \\
\hline
Fully-supervised & all & $12.4$ & $7.6$ & $23.2$\\
IMU-supervised & $3$ & $28.5 \pm 3.2$ & $21.3 \pm 2.7$ & $55.9 \pm 8.5$\\
IMU-supervised IR & $3$ & $20.1 \pm 2.9$ & $14.4 \pm 2.0$ &  $40.1 \pm 8.8$\\
\hline
\multicolumn{5}{c}{Warehouse dataset} \\
\hline
Fully-supervised & all & $7.2$ & $5.9$ & $15.3$\\
Channel Charting & $32$ & $30.9$ & $26.8$ & $108.5$\\
IMU-supervised & $1$ & $13.4 \pm 0.9$ & $11.5 \pm 0.7$ & $25.0 \pm 1.8$\\
IMU-supervised IR & $1$ & $9.4 \pm 1.0$ & $8.4 \pm 0.8$ & $17.2 \pm 2.3$\\
\hline
\end{tabular}
\label{tab:main_results}
\end{center}
\end{table}

\begin{table}[htbp]
\def\arraystretch{1.3}
\caption{IMU absolute error [cm]}
\begin{center}
\begin{tabular}{|c|c|c|c|}
\hline
Method & Mean & Median & 90\textsuperscript{th} perc. \\
\hline
Dead-reckoning & $110.9 \pm 10.7$ & $68.0 \pm 9.0$ & $285.3 \pm 25.4$ \\
Forward-backward & $17.8 \pm 1.2$ & $14.0 \pm 1.3$ & $41.0 \pm 3.4$ \\
\hline
\end{tabular}
\label{tab:ablation}
\end{center}
\end{table}

\subsection{Ablation studies}
The experiments in this section are performed on the Warehouse dataset.

\subsubsection{IMU error}
IMU pseudo-labels are imprecise, and we can measure their error with respect to precise ground-truth. Tab.~\ref{tab:ablation} includes two versions, basic dead-reckoning and integration with forward-backward correction. The dead-reckoning baseline demonstrates the IMU properties, and the corrected version shows the advantage of the correction algorithm. Additionally, we can see that the IMU-supervised model (Tab.~\ref{tab:main_results}) shows smaller error compared to the pseudo-labels. This suggests that the model was able to generalize reducing the error.

\subsubsection{K-nearest neighbors (k-NN)}
The problem of expensive position labels is not specific to neural networks. To demonstrate the compatibility of our solution with other positioning methods, we also evaluate a k-NN regressor as a possible positioning solution. In this case, for each test sample we look up 7 neighbors using $L_1$ distance for CSI features from train set and take the average pseudo-labeled position. The resulting mean absolute error is $14.8 \pm 1.1$ cm, which is close to NN result ($13.4 \pm 0.9$ cm). Thus, our pseudo-labels approach can be applied to different downstream positioning systems.

\begin{figure}[htbp]
\centerline{\includegraphics[width=7cm]{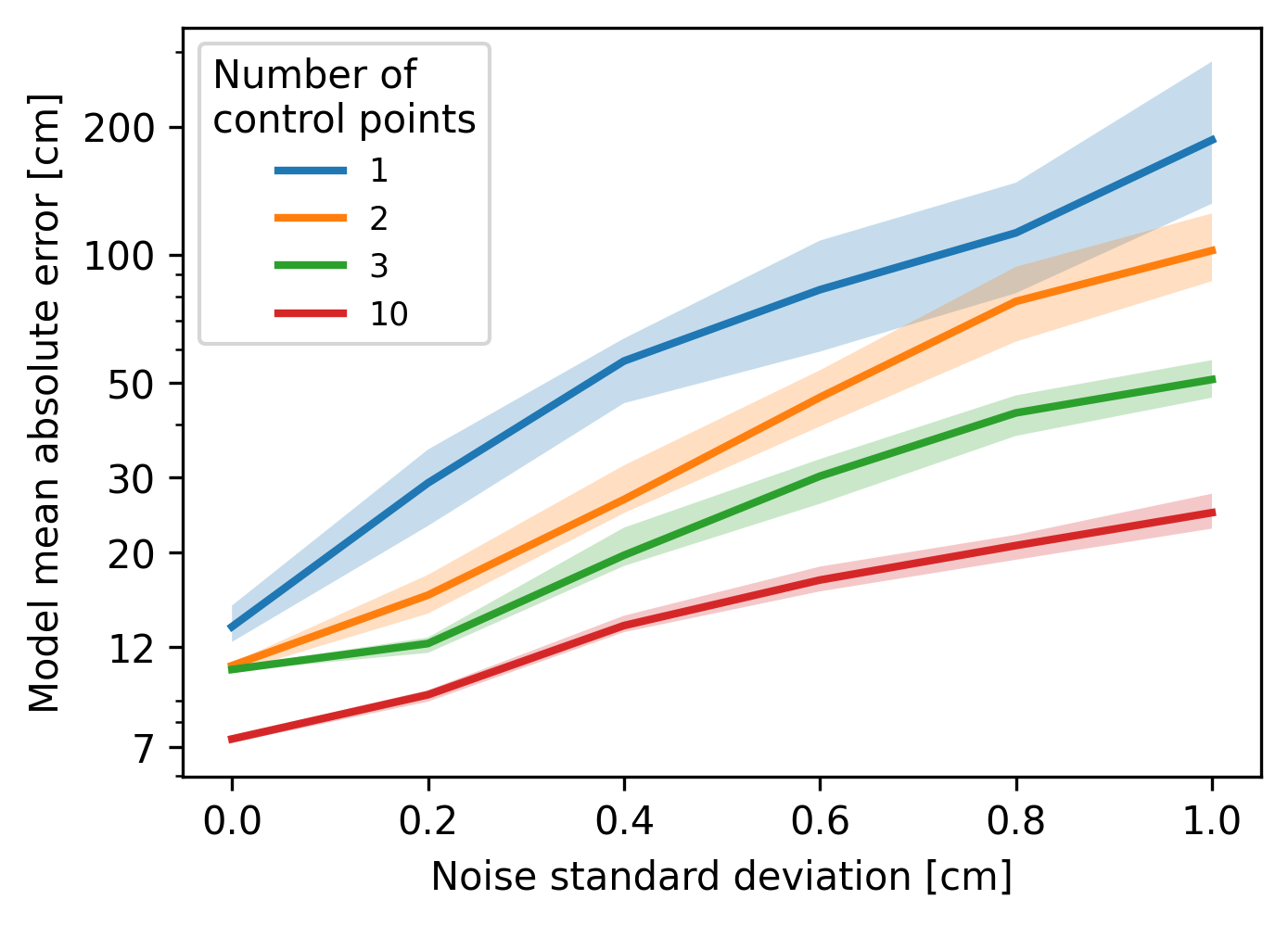}}
\caption{Dependency between IMU-supervised model accuracy and positions noise for Warehouse dataset. The light area is 0.1 and 0.9 quantiles, the line is median for 30 seeds. Vertical axis is in logarithmic scale.}
\label{fig:noise}
\end{figure}

\subsubsection{Imprecise position measurements}
We add Gaussian noise to positions to imitate noisy measurements in control points. Resulting positions and velocities are used as starting values for IMU integration. Fig.~\ref{fig:noise} shows that the accuracy degrades with noise, although additional control points from the other corners of the area help mitigate this effect.
Note that the velocity is calculated as a discrete difference of positions and is therefore heavily distorted by noise. In real applications, more sophisticated techniques can be used to obtain a more robust velocity approximation, such as the Kalman filter.

\subsubsection{Control points}
\label{sec:cpoints}
\begin{figure}[htbp!]
\centerline{\includegraphics[width=7cm]{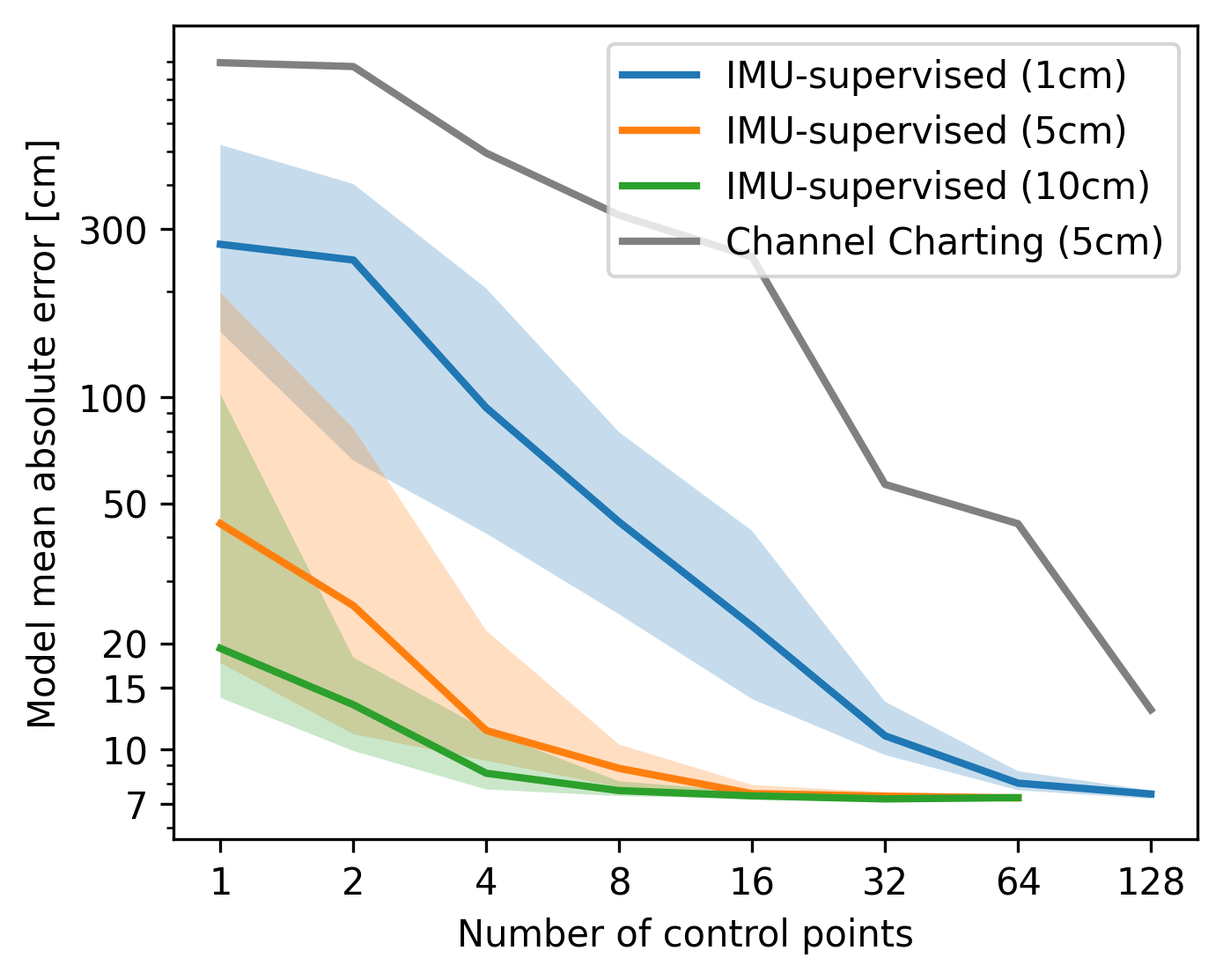}}
\caption{Dependency between model accuracy and the control points with different radius for Warehouse dataset. The light area is 0.1 and 0.9 quantiles, the line is the median for 30 seeds. Both axes are in logarithmic scale.}
\label{fig:cpoints}
\end{figure}

The choice of control points has significant impact on the final model accuracy. If there are few control points, the trajectory between them is too long and the IMU measurements accumulate the error.
In the main experiment, we use one control point at starting position with \SI{20}{\centi\metre} radius. In this case, our dataset is represented as 35 subtrajectories between control points with median length of \SI{51}{\metre}.
Additionally, we demonstrate the accuracy for random control points with radius \SI{1}{\centi\metre}, \SI{5}{\centi\metre} and \SI{10}{\centi\metre} on Fig.~\ref{fig:cpoints}.
Note that control points are spread uniformly, which is the best setting for Channel Charting method. Nevertheless, it can be seen that IMU supervision provides a much stronger signal.

\begin{table}[htbp]
\def\arraystretch{1}
\caption{imuSensor configuration}
\begin{center}
\begin{tabular}{|c|c|c|}
\hline
TemperatureScaleFactor & $0.008$ & \SI{}{\percent\per\celsius} \\
ConstantBias & $0.1962$ & $\SI{}{\metre\per\second\squared}$ \\
TemperatureBias & $0.0014715$ & $\SI{}{\metre\per\second\squared\per\celsius}$ \\
NoiseDensity & $0.0012361$ & $\SI{}{\metre\per\second\squared\per\sqrt{\hertz}}$ \\
\hline
\end{tabular}
\label{tab:imu_spec}
\end{center}
\end{table}

\section{Conclusion}
Localization is an important task, which raises several challenges. Often, specific equipment, such as Lidar, is not available. Moreover, common sources of information, such as visual observations, may not preserve privacy. The use of 5G signal indoor is widespread and CSI-based localization has emerged as a supervised learning task. In this work, we have addressed the bottleneck of such methods, the dependency on hard-to-acquire position labels. We have introduced the practical algorithm to obtain pseudo-labels from IMU measurements and confirmed our findings with empirical results. Our method demonstrates high accuracy (\SI{9.4}{\centi\metre} mean absolute error) for actual measurements.

\bibliography{bibliography}
\bibliographystyle{IEEEtran}
\end{document}